\documentstyle[12pt]{article}

\catcode`\@=11
\def\numberbysection{\@addtoreset{equation}{section}
        \def\theequation{\arabic{equation}}}
\def\beq{\begin{equation}}
\def\eeq{\end{equation}}
\def\half{\frac{1}{2}}
\numberbysection
\begin{document}
\begin{titlepage}
\begin{center}
\hfill CERN-TH / 95-193 \\
\vskip 1.in
{\Large
\bf Non-perturbative Particle Dynamics \\
in (2+1)-Gravity \footnote{Work supported in part by M.U.R.S.T., Italy.
\\ ${}^{**}$ On sabbatical leave of absence from Dipartimento di Fisica,
Universit\`a di Firenze and INFN, Sezione di Firenze, Italy.\\
{} \\
CERN-TH / 95-193 \\
July 1995 } }
\vskip 0.2in
A. Bellini \\
[.2in] {\em Dipartimento di Fisica, Universit\`a di Firenze, Italy,}
\vskip 0.2in
M. Ciafaloni${}^{**}$
\vskip 0.2in
{\em Theoretical Physics Division, CERN, Geneva}
 \vskip 0.2in  and  \vskip 0.2in
P. Valtancoli \\
\vskip 0.2in {\em Dipartimento di Fisica, Universit\`a di Firenze \\
and INFN, Sezione di Firenze, Italy} \\
\end{center}
\vskip .5in
\begin{abstract}
We construct a non-perturbative, single-valued solution for the metric and the
motion of two interacting particles in ($2+1$)-Gravity, by using a Coulomb
gauge
of conformal type. The method provides the mapping from multivalued
( minkowskian) coordinates to single-valued ones, which solves the non-abelian
monodromies due to particles' momenta and can be applied also to the general
N-body case.
\end{abstract}
\vfill
\end{titlepage}
\pagenumbering{arabic}
%
%
The classical \cite{a1}-\cite{Gott2} and quantum \cite{a3}-\cite{a5} structure
of $(2+1)$-Gravity coupled to matter has
been thoroughly investigated in the past by using locally Minkowskian
coordinates and/or its topological relation to Chern-Simons Poincare' gauge
theories \cite{a6}-\cite{a7}.

The choice of Minkowskian coordinates is possible because in ($2+1$)-dimensions
the space is flat outside the ( pointlike ) matter sources. However, the
localized curvature due to particles' momenta implies that the Minkowskian
coordinates are not single-valued, but are changed by a Poincare'
transformation
by parallel transport around the sources ( DJH matching conditions
\cite{a2} ). This implies that the metric description requires singularity
tails carried by each particle \cite{a8}-\cite{a9}.

On the other hand, in order to define the scattering problem, and in general
local particle properties, it is convenient to look for regular gauges, in
which the metric is not Minkowskian, but is single-valued and is singular
only at the particle sites. A method for constructing the coordinate
transformation from singular to regular gauges was given in Ref.
\cite{a8}, but an explicit solution was exhibited only in the massless limit
and in an Aichelburg-Sexl \cite{a8}-\cite{a10} gauge, of covariant type. In
the general massive case only partial perturbative results are available
\cite{a11}-\cite{a14}.

The purpose of this note is to propose an alternative non-perturbative method
to construct the above coordinate transformation, and thus the regular metric
for any number of particles, and to determine the main features of the two-body
problem in an "exact" way. A key ingredient of the present solution is our
choice of gauge \cite{a14}-\cite{a12}, which is of conformal type and is also
of Coulomb type \cite{a3}-\cite{a13} , i.e., it yields an instantaneous
propagation.

To set up the problem, let $X^a \equiv ( T / Z / {\bar Z} )$ denote the
Minkowskian coordinates and $x^\mu \equiv ( t / z / {\bar z} )$ the
single-valued ones. They are related by a dreibein $E^a_\mu = ( A^a, B^a
, \widetilde{B}^a )$ such that
\beq dX^a = E^a_\mu dx^\mu = A^a ( x ) dt + B^a (x) dz + \widetilde{B}^a ( x )
d{\bar z} , \label{1} \eeq
where the $A$' s and $B$' s are to be determined by conditions to be given
shortly.

Since the $X$'s are Minkowskian, the line element is given by
\beq  ds^2 = \eta_{ab} dX^a dX^b = g_{\mu\nu} dx^\mu dx^\nu \label{2} \eeq
and therefore the single-valued metric tensor $g_{\mu\nu}$ is obtained as
\beq g_{\mu\nu} = E^a_\mu E_{a\nu}  , \label{3}\eeq
where the $a$ indices are lowered by the Minkowskian metric $\eta_{ab}$,
with non-vanishing components $\eta_{00} = - 2 \eta_{z {\bar z} } = 1 $.

For the ( multivalued ) coordinates $X^a$ to exist , the dreibein must
satisfy the integrability conditions
\beq  \partial_{[\mu} E^a_{\nu]} = 0   \ \ \ \ ( \mu , \nu = 0 , z ,
{\bar z} ) . \label{4}\eeq
The latter hold in the region outside the singularity tails departing from
each particle source, which are needed in order to define a Riemann surface
for the $X$'s , and carry a non-trivial, localized spin connection, discussed
elsewhere \cite{a8}-\cite{a9}.

Following Ref. \cite{a14}, we choose to work in a generalized Coulomb gauge of
conformal type, which in the present first-order formalism is defined
by\footnote{
The condition (5) insures also the vanishing of the extrinsic curvature
$\Gamma_{0,z \bar z}=0$, and is therefore
equivalent to the instantaneous gauge obtained in Ref. [13] . }
\beq  \partial \cdot E^a = \partial_z E_{\bar z}^a + \partial_{\bar z} E^a_z =
0
\label{44} \eeq
\beq  g_{zz} = g_{{\bar z}{\bar z}} = 0. \label{5} \eeq
Because of Eqs. (\ref{4}) and (\ref{44}), the dreibein components satisfy the
conditions
\beq \partial_{\bar z} B^a =  \partial_z \widetilde{B}^a = 0 \label{55}\eeq
\begin{eqnarray}
\partial_z A^a & = & \partial_0 B^a ( z, t ),  \nonumber \\
\partial_{\bar z} A^a & = & \partial_0 \widetilde{B}^a ( {\bar z} , t ),
\label{6}\end{eqnarray}
Therefore, $B^a ( z,t) $ ( $\widetilde{B}^a ( {\bar z}, t ) ) $ are analytic
( anti-analytic ) functions and $A^a ( z, {\bar z} , t ) $ are harmonic
functions, i.e., sums of analytic and anti-analytic ones.

Furthermore, because of Eq. (\ref{5}) , $B^a$ and $\widetilde{B}^a$ are
null-vectors so
that, by using straightforward conjugation properties we can parametrize
\begin{eqnarray}
B^a & = & N ( z , t ) W^a ( z , t ) \ , \ \ \ \ \widetilde{B}^a = {\bar N}
( {\bar z} , t ) \widetilde{W}^a ( {\bar z}, t ) ,  \nonumber \\
W^a & \equiv &  {(f^{\prime})}^{-1} ( f / 1 / f^2 ) \ , \ \ \
\widetilde{W}^a \equiv {({\bar f}^{\prime})}^{-1} ( {\bar f}
/ {\bar f}^2 / 1 ) \ , \ \  \label{7}\end{eqnarray}
with $ W^2 = \widetilde{W}^2 = 0 $, and
\beq A^a = ( a / A / {\bar A} ) \ , \ \ \ a = {\bar a} , \label{8}\eeq
where $N(z,t)$, $f(z,t)$ and $f^{\prime} = df / dz $ are analytic functions,
and
$ a ( z, {\bar z}, t ), A ( z , {\bar z}, t )$ are harmonic ones.

It is now straightforward to obtain the components of the metric tensor
(\ref{3}) in
the form
\begin{eqnarray}
\qquad \qquad \qquad
- 2 g_{z {\bar z}} & \equiv & e^{2 \phi} = {\left| \frac{N}{f^{'}} \right|}^2
{ ( 1 - | f |^2 )}^2 = |N|^2 ( - 2 W \cdot \widetilde{W} ) ,
\ \ \ \nonumber \qquad \qquad \qquad \qquad (11a) \\ \qquad \qquad \qquad
g_{0z} & \equiv & \half {\bar \beta} e^{2\phi} = N W_a A^a \ \ , \ \ \
g_{0{\bar z}} \equiv \half \beta e^{2\phi} = {\bar N} \widetilde{W}_a A^a \ \ ,
\nonumber \ \! \qquad \qquad \qquad (11b) \\ \qquad \qquad \qquad
g_{00} & \equiv & \alpha^2 - {| \beta |}^2 e^{2 \phi } \ \ , \ \ \alpha
= V_a A^a , \ \nonumber \quad \qquad \qquad \qquad
\qquad \qquad \qquad \qquad (11c)
\label{9}\end{eqnarray}
\addtocounter{equation}{1}
where we have defined the vector
\beq V^a \equiv {( 1 - | f |^2 )}^{-1} ( 1 + | f |^2 \ / \ 2 {\bar f}
\ / \ 2 f ) = \epsilon^{a}_{bc} W^{b}\widetilde{W}^{c}
( W \cdot \widetilde{W} )^{-1} . \label{10}\eeq

Eqs. (11) and (\ref{10}) express the fields $ \alpha, \beta, \phi $
corresponding
to four real variables, in terms of the functions $ f , N , A , a $
corresponding to seven real variables. This is because the metric determines
the dreibein only up to local Lorentz transformations, in this case the
$3$-parameter $O(2,1)$ group.

If $N$, $f$ ( $a$ , $A$ ) are analytic ( harmonic ) everywhere in the
$z$-plane,
then Eq. (11) describes a pure gauge degree of freedom, satisfying
the Einstein equations with vanishing energy-momentum tensor, and we end up
with a truly Minkowskian geometry.

Particle sources with masses $m_i$ and Minkowskian momenta $P_i^a$ ( $i = 1,
..., N$ ) yield instead singularities of the dreibein at the particle sites
$z = \xi_i (t)$. They will be introduced in the following by the DJH matching
conditions \cite{a2}, i.e., by the requirement that
\beq  ( d X^a )_{II} =  {(L_i)}_b^a ( d X^b )_{I} \ , \ \  i = 1, ... , N ,
\label{11}\eeq
where
\beq L_i = exp ( i J_a P^a_i ) \ , \ \  {(J_a)}_{bc} = \epsilon_{abc} \ ,
\label{12}\eeq
denote the holonomies of the spin connection \footnote{The spin connection
is localized on the tails \cite{a8} , but its form will not be discussed here,
since we use the global property (\ref{11}).}, corresponding to loops around
the
singularity
of particle $i$, and labels I ( II ) denote determinations of the $X^a$
coordinates before ( after ) the loop operation.

The conditions (\ref{11}) imply that the dreibein components are multivalued
and
transform as $O(2,1)$ vectors under application of the $L_i$'s and their
products, so as to yield an invariant, i.e. , single-valued metric tensor
given by the explicitly scalar expressions in Eq. (11).

Suppose now we are able to find an analytic function $f(z,t)$, with branch
points at the particle sites $z = \xi_i (t)$ such that, when $z$ turns around
$\xi_i$, the $f$ transforms as a projective representation of the monodromies
(\ref{11}), i.e.,
\beq f(z,t) \rightarrow \frac{a_i f(z,t) + b_i}{{\bar b}_i f(z,t) + {\bar a}_i}
\
, \ \ i = 1, ... , N , \label{13}\eeq
where the $a$'s and $b$'s parametrize the spin $\half$ representations of the
loop transformations in (\ref{12}). Then the $W$'s , constructed out of $f$ in
Eq. (\ref{7}), transform as the adjoint ( vector ) representation of $O(2,1)$,
because they are obtained by applying the generators
\beq L^a = ( f \frac{\partial}{\partial f} / \frac{\partial}{\partial f} /
f^2 \frac{\partial}{\partial f} ) \label{14}\eeq
to the single-valued variable $z$. The $\widetilde{W}$'s do the same for the
equivalent conjugate representation. It follows that $N(z,t)$ must be
single-valued, and is at most meromorphic, with poles at $z=\xi_i$.

As for $A^a$, its vector transformation property is insured by the consistency
conditions in Eq. (\ref{6}), up to quadratures. Similarly, the vector
character of $V^a$ in (\ref{10}) under the transformation (\ref{13}) can be
checked  explicitly. As a
consequence, a solution to the conditions (\ref{13}) will automatically provide
the correct transformation properties of the dreibein and a single-valued
metric, and together with Eq. (\ref{6}) has a good chance of determining the
whole  problem.

The simplest example of condition (\ref{13}) is for one particle of mass $m$ at
rest.
In such case the loop transformation (\ref{12}) is a rotation of the deficit
angle
\beq 2\pi ( 1 - \alpha ) = m \ \ \ \ ( 8 \pi G = 1 ) , \label{15}\eeq
and Eq. (\ref{11}) is just multiplication by the corresponding phase factor
$exp ( im ) $. Therefore, for a particle at the origin,
\beq f(z,t) = K \ z^\lambda \ , \ \ \lambda = \frac{m}{2\pi} \ \ ( {\rm mod} \
n )
, \label{16}\eeq
where, however, the constant $K = O(V)$ should be considered as infinitesimal
with the velocity \footnote{This feature is shared by the general case to be
discussed below, and is rooted in the fact that the particles interaction
becomes trivial in the static limit, due to the lack of a Newtonian force.}
of the particle, so as to yield vanishing mixed components of the metric
in Eq. (11). In this limit, also $N \sim K / z$ vanishes, and the only finite
quantities are, up to a scale transformation and with $n=0$,
\beq \frac{f^{\prime}}{N} = \frac{1}{\alpha} z^{m / 2 \pi} \ , \ \
e^{2\phi} = \alpha^2 {| z |}^{ - m / \pi} = - 2 g_{z {\bar z}} ,
\label{17}\eeq
which yield the well-known \cite{a1}-\cite{a2} conical geometry in the
conformal gauge:
\begin{eqnarray}
ds^2 & = & dt^2 - \alpha^2 {|z|}^{- m / \pi} {| dz |}^2 = \nonumber \\
     & = & dT^2 - {| dZ |}^2 \ , \ \ ( arg Z < \pi \alpha ) .
\label{18}\end{eqnarray}

Next comes the two-body problem, with masses $m_1$ and $m_2$ , and momenta
\beq P_1 = ( E_1 / P / {\bar P} ) \ , \ \  P_2 = ( E_2 / - P / - {\bar P} )
\label{19}\eeq
in the Minkowskian c.m. frame. In terms of the rescaled variable
\beq \zeta ( z, t ) = \frac{z - \xi_1}{\xi_2 - \xi_1} = \frac{z - \xi_1}{\xi}
\ , \ \  \label{20}\eeq
the function $f(z,t)$ has now branch points at $\zeta = 0$ and $\zeta = 1$
( and $\zeta = \infty$ ), around which it has to transform as in Eq.
(\ref{13}),
with
\begin{eqnarray}
a_i & = & cos \ \frac{m_i}{2} \ + \ i \ \gamma_i \ sin \frac{m_i}{2}, \ \
b_i = - i \ \gamma_i \ {\bar V}_i \ sin \frac{m_i}{2} \ \ , \nonumber \\
V_{1,2} & \equiv & \pm \frac{P}{E_{1,2}} \ ,  \ \ \gamma_i \equiv
{( 1 - | V_i |^2 )}^{-\half} \ , \ \  i = 1, 2.
\label{21}\end{eqnarray}

The difficulty now lies in the fact that $L_1 , L_2 $ do not commute,
because of the relative speed, and thus cannot be brought together to the
form of a phase transformation. Nevertheless, we can use the analyticity
properties of the solution of second order differential equations around
Fuchsian singularities \cite{a15}
 in order to obtain $f(z,t)$ as the ratio of properly
chosen independent solutions with three singularities, i.e., essentially
hypergeometric functions.

Indeed, after some algebra, we find the expression
\beq f(z,t) = e^{-i \theta_V} \frac{ f_{(1)} (z,t) - th \half \eta_1}
{ 1 - th \half \eta_1 f_{(1)} (z,t) } \ , \label{22}\eeq
where
\beq f_{(1)} (z,t) \ = \ cth \half ( \eta_1 - \eta_2 ) \ \zeta^\lambda \
\frac{ \widetilde{F} ( \half ( 1 + \mu + \lambda - \nu ) , \half ( 1 - \mu
+ \lambda - \nu ) ; 1 + \lambda ; \zeta ) }
{ \widetilde{F} ( \half ( 1 + \mu - \lambda - \nu ) , \half ( 1 - \mu
- \lambda - \nu ) ; 1 - \lambda ; \zeta ) } \label{23}\eeq
has the meaning of $f$-function in the particle $1$ rest frame, $\eta_i =
th^{-1} V_i$ denote the velocity boosts, $\theta_V$ the relative velocity
phase, $ \widetilde{F} ( a, b, c ; z ) \equiv \Gamma (a) \Gamma (b) \Gamma^{-1}
(c) F ( a , b , c ; z ) $ is a modified hypergeometric function, and the
indices $\lambda$, $\nu$ and $\mu$ are related to the masses $m_1$ , $m_2$
and invariant mass $\cal M$ as follows
\begin{eqnarray}
\lambda & = & \pm \frac{m_1}{2\pi} \ ( {\rm mod}. \ n_1 ) \ , \ \
\nu = \pm \frac{m_2}{2\pi} \ ( {\rm mod}. n_2 ) ,  \nonumber \\
\mu & = & \pm \left( \frac{{\cal M}}{2\pi} - 1 \right) \ ( {\rm mod}. -n_1
-n_2 + 2n ) , \label{24}\end{eqnarray}
where ${\cal M}$, corresponding to the topological invariant $Tr ( L_1 L_2 )$,
is given by \cite{a7}
\beq cos \frac{{\cal M}}{2} = cos \frac{m_1}{2} cos \frac{m_2}{2} -
sin \frac{m_1}{2} sin \frac{m_2}{2} \ \frac{P_1 \cdot P_2}{m_1 m_2} .
\label{25}\eeq

The solutions (\ref{22}) and (\ref{23}) are obtained by observing
\cite{a16} that, if
$y_1$ and $y_2$ are independent solutions of the equation
\beq y^{\prime\prime} + \frac{1}{4} \left( \frac{1 - \lambda^2}{\zeta^2} +
\frac{1 - \nu^2}{{( 1 - \zeta )}^2} + \frac{1 - \lambda^2 - \nu^2 + \mu^2}
{\zeta ( 1 - \zeta )} \right) y = 0 , \label{26}\eeq
then $ f = y_1 / y_2 $ transforms according to a subgroup of $SL(2,C)$
around the branch points $\zeta = 0$ and $\zeta = 1$. By adjusting the $y$'s
and their indices to our $O(2,1)$ case in Eq. (\ref{21}), Eqs. (\ref{22}) and
(\ref{23}) follow. \\
Note that $f(z,t)$ is time-dependent only through the rescaled variable
$\zeta (z,t)$ in Eq.(\ref{20}), because the momenta $P_1$, $P_2$ are the
constants
of motion of our problem \cite{a8}-\cite{a14}. Note also that different
determinations of $f$
due to different choices of the integers $n_1$, $n_2$, $n$, correspond in
general to different behaviours close to the singularities and for $z
\rightarrow \infty$.
In the following, we shall choose the \ $+$ \  determination of signs and we
shall also
set $n_1 = n_2 = n = 0$, in order to match with perturbative results
\cite{a14}.

The complete determination of the meromorphic function $N(z,t)$ appears to
be harder. We shall exclude a holomorphic ( constant ) behaviour because
at least pole singularities are needed to build non-trivial sources. Assuming
simple poles ( corresponding to $\delta$-function energy-momentum density ),
we think
that the residues should be related in order to avoid zeros of the
determinant\footnote{We are indebted to Camillo Imbimbo for a discussion on
this point.
The presence of zeros could also be cancelled by additional
spurious singularities in $f(z,t)$. We are assuming here a minimal set related
to the particle sites, which appears appropriate in the c.m. system.}
\beq \sqrt{|g|} = | E | = \alpha e^{2\phi} = \left| \frac{N}{f^{'}} \right|^2
{( 1 - | f |^2 )}^2 ( V_a A^a ) . \label{27}\eeq

Therefore, we shall take the ansatz
\beq N(z,t) = \frac{R(\xi(t)))}{( z - \xi_1)(\xi_2 - z)} =
\frac{R(\xi(t))}{\xi(t)^2} \frac{1}{\zeta ( 1 - \zeta )} , \label{28}\eeq
where $\xi(t) \equiv \xi_2 (t) - \xi_1 (t)$. A form of type (\ref{28}) checks
also with perturbative results \cite{a11}-\cite{a14}.

We are now in a position to discuss the functional relation of the
coordinates $X^a$ and $x^\mu$ implied by Eqs. (\ref{1}), (\ref{6}),
(\ref{7}) and by Eqs. (\ref{23}) and (\ref{28}). By integrating Eq.
(\ref{1}) out of particle $1$, say, we obtain
\beq X^a = X^a_1 (t) + \int^z_{\xi_1} \ dz \ N W^a ( z, t ) +
\int^{\bar z}_{{\bar \xi}_1} \ d{\bar z} \ {\bar N} \widetilde{W}^a ( {\bar z},
t ) , \label{29}\eeq
where we denote the Minkowskian $1$-trajectory by
\beq X_1^a (t) = B_1^a + V_1^a T_1 (t) \ \ \ ( V_1^a \equiv P_1^a / E_1 ).
\label{30}\eeq
By then inserting the ansatz (\ref{28}) into Eq. (\ref{29}) we obtain
\beq X^a = B_1^a + V_1^a T_1 + R( \xi(t)) \int^{\zeta(z,t)}_0 \
\frac{d\zeta}{\zeta ( 1 - \zeta )} W^a ( \zeta ) +
{\bar R}( {\bar \xi}(t)) \int^{{\bar \zeta}(z,t)}_0 \
\frac{d{\bar \zeta}}{{\bar \zeta} ( 1 - {\bar \zeta} )} \widetilde{W}^a ( {\bar
\zeta} ) ,\label{31}\eeq
and, by a time-derivative,
\beq A^a = V_1^a \dot{T}_1 + \partial_t \left( R ( \xi (t)) \ I^a ( 0,
\zeta (z,t) ) + {\bar R} ( {\bar \xi} (t))
\widetilde{I}^a ( 0 , {\bar \zeta} ) \right) , \label{32}\eeq
where we have introduced the notation
\beq I^a ( 0 , \zeta ) = \int^\zeta_0 \frac{d\zeta}{\zeta ( 1 - \zeta )}
W^a ( \zeta ) \label{33}\eeq
and a similar one for $\widetilde{W}$.

The expression for A in Eq. (\ref{32}) satisfies the consistency condition
(\ref{6})
by construction, and the monodromy vector transformation by inspection.
Furthermore, the $z$-dependence in Eqs. (\ref{31}) and (\ref{32}) is embodied
in the
integrals $I^a ( 0 , \zeta ( z, t ))$, which in turn are determined by the
functional form of $f(\zeta)$ in Eqs. (\ref{22}) and (\ref{23}).

So far, the time-dependent residue function $R( \xi (t) )$ in Eq.
(\ref{32})
appears undetermined and so is, therefore, the relative motion trajectory
$\xi (t)$. In fact, we have still to insure that we are not in a rotating
frame at space infinity or , in other words, that the affine connection
vanishes fast enough asymptotically. This asymptotic condition implies that
$A^a ( t, z , {\bar z} )$ should be at most logarithmic, for large $|z|$, and
therefore by Eq. (\ref{32}), that
\beq R ( \xi ) \  \zeta \partial_{\zeta} I^a ( 0, \zeta ) \rightarrow
I^a ( 0 , \zeta ) \ \xi \partial_{\xi} R ( \xi ) \ \ , \ \
\left( |\zeta| \simeq \left| \frac{ z }{\xi } \right| \gg 1 \right).
\label{34}\eeq

On the other hand, it is easy to check that, by Eqs. (\ref{22}) and
(\ref{23}), $I^a$
increases as
\beq I^a ( 0 , \zeta ) \simeq C^a \zeta^{1 - {\cal M} / 2\pi } \ \ ,
| \zeta | \gg 1 , \ \ 0 < {\cal M} < 2 \pi , \label{35}\eeq
where $\cal M$ is the invariant mass (\ref{25}). Therefore, by Eq. (\ref{34})
we must require
\beq R( \xi(t)) = C {( \xi(t) )}^{1 - \frac{{\cal M}}{2\pi}} \ \,
\label{36}\eeq
which determines $R$ up to a scale factor, and thus $N$, $A^a$ and the
metric as functions of $\zeta ( z, t)$, $\xi (t)$ and of the constants of
motion.

Finally, we have still to use the equations of motion for particle $2$, which
in integrated form read, by Eq. (\ref{31}),
\beq B_2^a - B_1^a + T_2 V_2^a - T_1 V_1^a = C \xi^{1- {\cal M} / 2\pi }
I^a ( 0, 1 ) + {\bar C} {\bar \xi}^{1-  {\cal M} / 2\pi }
\widetilde{I}^a ( 0, 1 ). \label{37}\eeq

Since $I^a$ and $\widetilde{I}^a$ are calculable constants, functions of
$P_1^a$
and $P_2^a$, Eq. (\ref{37}) determines the relative time variable $T_1 (t) -
T_2 (t)$
and the trajectory $\xi (t)$ up to an overall time reparametrization and a
scale freedom provided by $C$.

Without discussing the solution of Eq. (\ref{37}) in detail, it is rather
clear that
$\xi^{1- {\cal M} / 2\pi}$ should have, for large times, the same phase as
$ ( V_1 - V_2 ) t + i B $, where $B$ is the relative impact parameter.
It follows that $\xi$ should rotate by $\pi
{( 1 - {\cal M} / 2\pi )}^{-1}$ as time varies from $- \infty$ to
$+ \infty$, and that the corresponding scattering angle is
\beq \Theta ( {\cal M} ) = \frac{{\cal M}}{2} {( 1 -
\frac{{\cal M}}{2\pi} )}^{-1}
\label{38}\eeq
consistently with an early suggestion by 't Hooft \cite{a3}.

We have so far analyzed in detail the two-body case. However, the method just
outlined applies in general to $N$ particles, provided we are able to solve the
monodromies (\ref{11}) by the auxiliary function $f$, transforming as in
Eq. (\ref{13}).

For $N$ particles, we expect that the corresponding second-order differential
equation should have at least $N+1$ regular singularities, one of which at
infinity.
Since only the difference of indices, related to physical masses, matters for
the branch point
behaviour of $f$, it seems that $N+1$ singularities are not enough for $ N > 2
$:
they provide $ 2 N - 1 $ parameters, instead of the $ 3 N - 3 $ which are
needed
( $N$ three-momenta with $O(2,1)$ invariance ). Hence, for $ N > 2 $, some
extra
singularities are expected in the Schwartzian derivative [17] of $f$, which are
not
branch points of $f$, but rather zeros of $f'$ .

Several comments are in order. First of all, the basic simplification which
allows to deal with the monodromy properties in a single complex plane is
rooted
in the $3$-dimensional nature of the problem, according to which the Coulomb
condition in Eq. (\ref{44}) implies the analyticity ( harmonicity ) properties
in
Eq. (\ref{55}) ( Eq. (\ref{6})). This in turn is equivalent to the
instantaneous propagation
in a second-order formalism \cite{a14}, and is due to the absence of physical
(transverse) gravitons. For this reason the time-evolution is coupled to the
$z$ dependence only through the rescaled variable $\zeta (z,t)$.

Secondly, the general method above can be explicitly checked by the
perturbative
calculations available for (i) first non-trivial order in $V_i$ and all-orders
in $G$ \cite{a11} and (ii) second-order in $G$ and any speed
\cite{a14}. For instance, in the
first case we find, from Ref. \cite{a11},
\begin{eqnarray}
 f(z,t) & = & \half ( {\bar V_1} - {\bar V_2} ) \int_0^{\zeta} dt
\ t^{m_1 / 2 \pi - 1} {( 1 - t )}^{m_2 / 2 \pi - 1} \
B^{-1} ( \frac{m_1}{2\pi} , \frac{m_2}{2\pi} ) - \half {\bar V_1} ,
\nonumber \\
 R( \xi ) & = & \half ( {\bar V_1} - {\bar V_2} )
B^{-1} ( \frac{m_1}{2\pi} , \frac{m_2}{2\pi} ) \xi^{1- \frac{m_1+m_2}{2\pi}} ,
\label{39}\end{eqnarray}
and explicit expressions, involving  hypergeometric functions, for $A$ and
$a$. The calculable parameters of Eq. (\ref{37}) are in this case, to first
order in $V_i$'s ,
\begin{eqnarray}
C I^a ( 0,1) & = & \ \left( i \ \frac{V_1 - V_2 }{2} \ J - i \
\frac{V_1}{2} I \ / \ I \ / \ 0  \right) , \ \ \
I = B \left( 1 - \frac{m_1}{2\pi} , 1 - \frac{m_2}{2\pi} \right) \ , \ \
\nonumber \\
J & = & B^{-1} \left( \frac{m_1}{2\pi} , \frac{m_2}{2\pi} \right)
\frac{ \psi ( 1 - m_2 / 2\pi ) - \psi ( m_1 / 2\pi ) }
{ 1 - m_1 / 2\pi - m_2 / 2\pi }.
\label{40}\end{eqnarray}
The check above still leaves open the question about the classification and the
physical meaning of alternative determinations for $f(z,t)$, corresponding to
different choices of $n$'s in Eq. (\ref{24}). This gives rise to different
behaviours around the singularity points, and in particular $z = \infty$,
around which asymptotic conditions of type in Eq. (\ref{34}) seem to work only
in a limited mass range ( e.g. , $0 < {\cal M } < 2 \pi $ ).

In addition, one should mention the possible ( non-perturbative ) zero of
the determinant (\ref{27}), which could occur at $|f| = 1$. With our choice of
indices one can show, by Eq.(\ref{23}), that
$| f_{(1)} ( \infty ) | , | f_{(1)} (1) | < 1$
in the naive mass range $ 0 < m_1, m_2, {\cal M} < 2 \pi$, hence  by the
maximum modulus theorems \cite{a17}
$ | f_{(1)} (z) | < 1 $
on the first sheet of the cut $z$-plane, and thus $|f(z)| < 1$ on any sheet ,
because Eq. (\ref{13}) preserves this inequality. Therefore, for proper values
of the masses, there are no problems with our choice.

On the other hand, if $ P_1 \cdot P_2 $ exceeds some critical value, the
invariant mass takes the form ${\cal M} =
2 \pi + i \sigma \ ( \ cos ({\cal M}/2) < - 1 ) $ and closed timelike curves
appear \cite{Gott}. In the same situation, since
\beq
{f_{(1)} (\infty)}^2  = \frac{sin \ \frac{1}{4} \ ( {\cal M} - m_1 - m_2 )}{sin
\ \frac{1}{4} \ ( {\cal M} + m_1 + m_2 )}
\frac{sin \ \frac{1}{4} \ ( {\cal M} - m_1 + m_2 )}{sin \ \frac{1}{4} \ ( {\cal
M} + m_1 - m_2 )}  \ , \label{43} \eeq
it is easy to realize that ${| f ( \infty ) |}^2 = 1$, and the gauge choice may
become pathological. Thus the restriction to
$ cos ({\cal M} / 2 ) > - 1 $ is natural, and also avoids CTC's.

Finally, let us remark that in our gauge, because of the instantaneous
propagation, the particles interact at all times, making decoupling properties
rather difficult to handle. For instance, comparing the expression (\ref{23})
with Eq. (\ref{16}), we see that only in the regions $ | z - \xi_1 | \ll | \xi
|
\ \ ( | z - \xi_2 | \ll | \xi | ) $ does the interacting metric look like the
single particle ones. In all other regions they considerably differ, at all
times.

This feature is to be contrasted with what happens in covariant-type gauges
\cite{a8}
in which the metric decouples in two single-particle ones at large times.
In particular, in the present case, the massless limit, which exists with some
care,
does not correspond to shock-wave scattering of Aichelburg-Sexl type.

As a consequence, the local space-time properties, and thus the scattering
angle in Eq. (\ref{38}), appear to be different than the ones in covariant
gauges \cite{a8}.
Whether this fact is to be related with the lack of true asymptotic decoupling
\cite{a18} in this gauge is an interesting question still to be investigated.

{\bf Acknowledgements }

It is a pleasure to thank Andrea Cappelli, Camillo Imbimbo, Giorgio Longhi,
Pietro Menotti and Gabriele Veneziano for interesting discussions and
suggestions.



\end{document}